# Structural, Electrical, Magnetic and Impedance Behaviour of $NdFeO_3$ Modified $Ba_{0.7}Sr_{0.3}TiO_3$ Ceramics

Anumeet Kaur, Lakhwant Singh*

Department of Physics, Guru Nanak Dev University, Amritsar

*Email (Corresponding Author): lakhwant@yahoo.com

## ABSTRACT

$NdFeO_3$ modified polycrystalline ceramics with composition $(NdFeO_3)_{0.1}$-$(Ba_{0.7}Sr_{0.3}TiO_3)_{0.9}$ (NFBST) ceramics have been synthesized via solid-state reaction route. The Rietveld refinement of the XRD data confirmed the existence of a tetragonal phase (P4mm) and a hexagonal phase (P63/mmc) in the prepared sample. The coexistence of phases has been also further confirmed from the Raman spectroscopy. The SEM image revealed dense microstructure with well packed grains of different sizes. At room temperature lossy P-E loop and weak ferromagnetism is observed in NFBST system. Complex impedance spectroscopy (CIS) as a function of frequency (100 Hz to 1 MHz) at different temperatures (RT to 700K) has been employed to study the electrical behaviour in NFBST ceramic. Two semicircular arcs in the Cole-Cole plot manifested the grain and grain boundary contribution in overall impedance. The detail complex modulus analysis and ac fitted conductivity data authenticated correlated barrier hopping CBH to be responsible for electrical as well as conduction phenomena in NFBST. Oxygen vacancies are responsible for relaxation and conduction processes in NFBST ceramics as divulged from the activation energy values estimated from electrical impedance, modulus, and conductivity data.

## 1. Introduction

The simple structure along with interesting electrical, magnetic, electrochemical, optical and catalytic properties of perovskite Ceramics has always got the significant attention of scientific community for their potential use in microelectronics, memories, resonators and energy storage devices etc [1,2]. Most of perovskite ceramics that are counted in the category of dominant ferroelectric ceramics are lead based such as Lead titanate (PT), Lead zirconate (PZT), Lead Strontium Titanate (PST) etc, as they show extraordinary dielectric properties.

Despite of their remarkable properties, they are being replaced by lead free materials as lead being highly toxic causes various environmental problems and health risks [3,4]. The extensively studied lead free alternative material is Barium titanate; which exhibits perovskite structure and has always been the first choice of researcher's because of its high dielectric constant, excellent piezoelectric and pyroelectric properties and large ferroelectric transition temperature around 400 K [5–7].

Barium strontium titanate ($Ba_{1-x}Sr_xTiO_3$; BST) with perovskite structure ($ABO_3$), is desired ferroelectric material that is extensively used as capacitors and DRAMS due to it's excellent properties such as high dielectric constant, low leakage current and tunable Curie temperature ($T_c$) [4,8]. Various compositions of BST have very high dielectric constant but the loss factors being high enough restricts it's practical applications [9]. However, to overcome this problem researchers have doped BST by various dopants i.e. with rare earth ions ($La^{3+}$, $Sm^{3+}$ etc.) and transition metal ions ($Co^{2+}$, $Ni^{2+}$, $Mn^{3+}$, $Fe^{2+}$, $Fe^{3+}$) [10,11]. In our previous report on Fe doped $Ba_{0.7}Sr_{0.3}TiO_3$ ceramics, we observed that incorporation of magnetic ion (Fe) resulted in high dielectric loss due to the formation of doubly ion oxygen vacancies and thereby destroying their electrical properties [12]. On the other hand, the temperature dependant magnetic studies showed that these Fe doped BST ceramics exhibited paramagnetic behavior at room temperature and even in the low temperature regime upto 2K [4]. Various reports are available in which issue of leakage current and high dielectric loss has been addressed by carrying out A-site lanthanide substitution using rare earth ions such as $Nd^{3+}$, $Gd^{3+}$ etc [13,14].

Moreover, the rare earth orthoferrites ($RFeO_3$) are studied for their unique magnetic, dielectric, magnetodielectric and multiferroic properties. The superexchange interactions between two different types of magnetic sublattices i.e $Fe^{3+}$ and $R^{3+}$ results in interesting magnetic properties [15–17]. However, the literature survey shows that $RFeO_3$-$ABO_3$ based multiferroic composites such as $NdFeO_3$–$PbTiO_3$, $LaFeO_3$–$PbTiO_3$, $BiFeO_3$–$SrTiO_3$, $BiFeO_3$–$BaTiO_3$ and $NdFeO_3$–$SrTiO_3$ etc show enhancement in both ferroelectric and magnetic properties [18–20].

In view of this, we have chosen Neodymium ferrite ($NdFeO_3$), which has distorted orthorhombic perovskite structure having Pbnm space group with antiferromagnetic ordering, $T_N$ ~ 760 K [19]. As discussed above, it is expected that the $NdFeO_3$ doping in $Ba_{0.7}Sr_{0.3}TiO_3$ matrix will not only address the issue of high dielectric loss but will also help in modifying the magnetic behavior of BST ceramics. So, we here report on the structural, magnetic and electrical behavior of $(NdFeO_3)_{0.1}$-$(Ba_{0.7}Sr_{0.3}TiO_3)_{0.9}$ ceramics where $Nd^{3+}$ and $Fe^{3+}$ ions will be substituted at the A-site and B-Site in BST ($ABO_3$) system respectively.

## 2. Experimental

The bulk samples of $(NdFeO_3)_{0.1}$-$(Ba_{0.7}Sr_{0.3}TiO_3)_{0.9}$ (thereafter abbreviated as NFBST) ceramics were synthesized by conventional solid state reaction route. Raw materials ($BaCO_3$, $SrCO_3$, $Nd_2O_3$, $Fe_2O_3$ and $TiO_2$) were weighed in desired stoichiometric ratio and ball milled for 24 hours. For proper phase formation, calcination was done at 1000°C for 12 hrs followed by mixing PVA (Polyvinyl Alcohol). Pressing of PVA mixed powder in form of pellets was done using hydraulic press by applying pressure of 10 MPa. Sintering of prepared pellets was carried out at 1250 °C for 2 hrs for proper grain growth and densification of samples. X-ray Diffraction (XRD) at room temperature was carried out using a SHIMADZU (CuKα radiation, MAXima XRD -7000) to confirm the phase purity of NFBST samples. The Rietveld refinement of the XRD data was carried out using FullProf software. Scanning Electron Microscope (SEM) (Carl ZEISS Supra 55) was used to study the surface morphology. Density measurements were carried out using lab made setup based on Archimedes principle. Vibrating sample magnetometer (VSM) (Microsense E29) was used for performing magnetic measurements (M-H loop) at room temperature. For the dielectric and ferroelectric measurements sintered pellet of 10 mm diameter was polished, electroded (with silver paste) and further dried at 600 °C for 1 hr. For room temperature P-E loops measurement, the NFBST pellet was dipped in silicon oil to prevent the electrical arc appearance. The hysteresis loop was recorded with an automatic ferroelectric loop tracer based on Sawyer-Tower circuit (Marine India) at a frequency of 20 Hz and optimal field of 40 kV/cm. The dielectric data was collected as a function of frequency (100Hz–1MHz) in a temperature range from room temperature (RT) to 700 K Using NOVO-CONTROL (Alpha-A) high-performance frequency analyzer.

## 3. Results and Discussions

### 3.1 Phase Determination

For determining the phase of NFBST sample, X-Ray Diffractogram recorded at room temperature in the range from 20° to 80° at the scan speed of 1°/min is shown in Figure 1. The first closer observation at the XRD pattern clearly shows the presence of sharp and high intensity peaks which assure that the sample is well crystallised. All the Bragg reflections present in the powder X-ray diffraction pattern confirmed the phase purity and could be indexed as main perovskite reflections. In our previous work, we have reported that

$Ba_{0.7}Sr_{0.3}Fe_xTi_{(1-x)}O_3$ (where x=0,0.1) samples have a coexistence of Tetragonal and Cubic phase [4]. So, in the beginning Rietveld refinement for NFBST sample was attempted using phase coexistence model with tetragonal (T) (P4mm) and Cubic (C) (Pm-3m) phase which has been shown in Figure 1. Peak profiles were described by Thomson-Cox-Hasting Pseudo-Voigt function, while a sixth order polynomial was taken to fit the background. Occupancy parameters of all the ions were kept fixed at their nominal composition. Zero correction, scale factor, lattice parameters and isotropic thermal parameters ($B_{iso}$) were varied during the refinement. The mixed phase model (T+C) gave final $\chi^2$ value of 2.16. It is clear from the inset of Fig. 1(a) that there is significant mismatch between measured and calculated data (marked by arrow). So it was anticipated that this misfit may be due to the co-existence of some other phases. It has also been reported by various authors that the addition of Fe in $BaTiO_3$ lattice also forms hexagonal Phase [21–24]. So, a two phase model of a tetragonal phase (P4mm) and a hexagonal phase (P63/mmc) (T+H) was used as refinement strategy which lead to the improvement in the fit by reducing $\chi^2$ to 1.53. The refined lattice parameters, GOF ($\chi^2$), $B_{iso}$, R-factors and cell volume calculated from Rietveld refinement are given in Table 1. The fit between the observed and calculated profiles is satisfactory (inset Figure 1 (b)) and indicates the correctness of phase coexistence model.

**3.2 Raman Spectroscopy**

The co-existence of phases in NFBST system was further confirmed with the help of Raman spectroscopy as it is widely employed to understand the lattice distortions and crystallographic defects in the solids. It is well reported in literature that Raman active bonds characterizing the tetragonal phase of $BaTiO_3$ has sharp bands around 250 cm$^{-1}$ A1 (2TO) and 307 cm$^{-1}$ [B1, E (3TO + 2LO)] and the wide bands around 515 cm$^{-1}$ [A1 (3TO), E (4TO)] and 720 cm$^{-1}$ [A1 (3LO), E (4LO)] [22,23,25–28]. Figure 2 shows Raman spectra of NFBST sample along with BST ($Ba_{0.7}Sr_{0.3}TiO_3$) and BSTF ($Ba_{0.7}Sr_{0.3}Fe_{0.1}Ti_{0.9}O_3$). The BST and BSTF samples exhibit mixed phase of Tetragonal (T) and Cubic (C) as reported in our previous work. The BST Spectra clearly shows that with doping of Sr in $BaTiO_3$, the intensity of band at 728 cm$^{-1}$ decreased. An obvious slight frequency range shift and decrease in peak intensity is always observed in Raman spectra with doping in the host material. This decrease in intensity of 728 cm$^{-1}$ band can be related to structural changes occurring in BST i.e. an existence of cubic phase along with tetragonal. However with introduction of Fe in BST, we observe that the band at 236 cm$^{-1}$ and 306 cm$^{-1}$ almost merged into one and intensity of band at 512 cm$^{-1}$ gets reduced confirming decrease in the tetragonality, matches well with the XRD data. An appearance of a new peak in BSTF at 664 cm$^{-1}$ can be attributed to strained lattice transpired out of doping Fe at the Ti site [22].

The Raman spectrum of NFBST having distinguishable bands is shown in figure 2(a). The spectra clearly show new peaks and bands cropping up with the introduction of NdFeO$_3$ in Ba$_{0.7}$Sr$_{0.3}$TiO$_3$ matrix. Figure 2(b) shows the fitting of Raman data done using a set of Lorentzian peak shape in the wave number range of 70-900 cm$^{-1}$. A set of deconvoluted peaks at 118 cm$^{-1}$, 179 cm$^{-1}$ and 242 cm$^{-1}$ can be assigned to Eg, [A1(LO)] and [A1(TO)] mode respectively [29,30]. It has been reported that the two Raman bands at 303 cm$^{-1}$ and 738 cm$^{-1}$ are the characteristic bands of tetragonal structure and have mixed character of B1 and E modes that are derived from F2u cubic mode and also peak at 506 cm$^{-1}$ is a characteristic of cubic phase [31–33]. Thus it signifies that although the XRD Data has been fitted using tetragonal phase, but still the lattice parameters obtained after Reitveld refinement being almost comparable cannot rule out the presence of pseudo cubic phase. Moreover, the Raman data also exhibits a prominent peak at 641 cm$^{-1}$ which is attributed to the presence of hexagonal phase in the sample [22]. So, the results obtained from Raman data are in completely compatible with XRD analysis which confirms the presence of mixed phase in NFBST sample.

### 3.3 Scanning Electron Microscopy

Figure 3 displays the image of grain structure for NFBST ceramic recorded at magnification of 5KX and EHT of 15kV using Scanning Electron Microscope (SEM). It was observed that BSTF ceramic had granular microstructure with irregular shapes and some degree of agglomeration[12]. It is clearly seen that the NFBST sample is highly dense and has well packed grains of different sizes. Thus, it can be remarked that doping of NdFeO$_3$ in BST matrix has resulted in dense microstructure. However, density of the NFBST sample (calculated by lab made setup based on Archimedes Principle) is estimated to be 6.329 g/cm$^3$. As expected during sintering process i.e. at high temperatures, thermal energy endorses growth of grain boundary over the porous regions which subsequently reduces the pore volume and further densifies the concerned materials. An uniform grain growth is expected when the driving force is homogeneous at the grain boundary and non-uniform growth for the case of inhomogenous driving force which is having explicit dependence upon sintering temperature, diffusivity of individual grains, porosity etc [34].

## 3.4 Ferroelectric and Magnetic Properties

The room temperature P-E loop for NFBST sample recorded at the field of 40 kV has been shown in Figure 4 (a). We can clearly see that the observed P-E loop is not a typical saturated one however, it can vaguely be proclaimed as a lossy loop. The primary condition to have saturated ferroelectric loop is non-centrosymmetry. The XRD analysis shows that crystal structure of NFBST is a mixture of pseudo cubic and hexagonal, both being centrosymmetric do not favour ferroelectricity and thus have resulted in lossy P-E loop. Incorporation of acceptor type impurities like Fe, further generates positively charged oxygen vacancies for maintaining overall charge neutrality and as a consequence defect dipoles are created which behaves like elastic ones for having prominent ionic radii difference between $Ti^{4+}$ and $Fe^{3+}$ ions [35]. When an external electric field is applied, the torque will try to align the aforementioned elastic dipoles along the polarization direction and subsequently reduces their potential energy. The defects present around domain wall affix the polarization direction which becomes more complex during motion and subsequently lossy P-E hysteresis loops are observed [36].

The recorded magnetization versus magnetic field behavior (*M-H* loop) at room temperature of NFBST in the field $-15kOe \leq H \geq 15kOe$ is shown in Figure 4 (b). It is observed that the NFBST ceramic exhibits weak ferromagnetism and the value of remnant magnetization ($M_r$) is 1.48 (*$10^{-4}$ emu/g). The inset in the Figure 4 (b) shows the enlarged view of the *M-H* loop. Generally, the materials of the first transition series which have free atoms and ions with incomplete d shells posses net magnetic moment and consequently result in ferromagnetic behavior. Five unpaired electrons in 3d shell of $Fe^{3+}$ ion form an antiferromagnetic G-type structure and results in weak ferromagnetic character at room temperature [19]. On the other hand, such weak ferromagnetism in ceramic samples can also be understood on the basis of *F*-center exchange mechanism [37–39]. The existence of magnetic behaviour inside NFBST can also be anticipated due to the presence of $Fe^{3+}$- $V_O^{2-}$- $Fe^{3+}$ groups in the structure. An electron not only constitutes F-Center by getting trapped inside oxygen vacancy but also occupies an orbital which performs d-shells overlapping between two neighbouring iron. For $Fe^{3+}$, $3d^5$ orbital exists with unoccupied minority spin orbital. The exchange interaction comes into the picture via F-center as trapped electron has down spin (↓) and nearby two iron has up spin (↑) which will lead to the generation of ferromagnetic coupling. Here F center executes an identical role as bound magnetic polaron.

## 3.6 Dielectric Permittivity Analysis

In order to understand the effect of $NdFeO_3$ doping in $Ba_{0.70}Sr_{0.30}TiO_3$ (BST) ceramics, the temperature dependence of dielectric constant (ε′) for NFBST sample in the temperature range of 300K-700K is illustrated in Figure 5. The ε′ versus T plot usually represents the ferroelectric-paraelectric phase transition. It has been reported that BST and BSTF has phase transition at ~308 K and ~290 K respectively. We observe that with $NdFeO_3$ addition, no peak corresponding to phase transition is seen in 300K-700K. It has been reported that $Nd^{3+}$ addition up to x = 0.10 in $BaTiO_3$ downshifts the phase transition peak from 128°C to -25°C [40]. So, we also recorded the temperature dependence of dielectric constant (ε′) in lower temperature regime i.e. from 100K to 400K (shown in inset of Figure 3) but could not observe any peak corresponding to phase transition. Thus, it can be anticipated that for NFBST sample the phase transition exists even below 100K. However, it is also observed that the subsequent monotonous increase of dielectric permittivity ε′ at higher temperature is due to the rise of conductivity of the sample polarization resulting from the mobility of ions and defects in the material. The broad peaks of the dielectric permittivity are getting suppressed and also shifting to higher temperature with rise in frequency, which is a typical signature of thermally activated relaxation process occurring in the material [41].

Figure 6 also displays the dielectric loss (tan$\delta$) variation with temperature. The dielectric loss is almost invariant up to 500K and starts rising afterwards. At higher temperatures, the oxidation state of Fe ion changes spontaneously from $Fe^{3+}/Fe^{2+}$ resulting in high conductivity and leakage in the sample. The oxygen vacancies so formed while change in oxidation states maintains electrical neutrality, thereby giving rise to thermally activated hopping conduction. Moreover in ceramics, the dielectric losses also occur due to defects associated with grain boundaries[42]. The NFBST sample has dielectric loss value of nearly 0.01 for 1KHz frequency at room temperature whereas the dielectric loss in $Ba_{0.7}Sr_{0.3}TiO_3$ with 0.1% of Fe in our previous report was nearly around 0.5 [12]. So, it can be clearly observed that the doping of rare earth ion i.e Nd significantly reduced the dielectric losses in the ceramics.

Figure 7 shows that dielectric constant ε′ and dielectric loss ε″ at different temperatures (573K-648K) decreases with increasing frequency. This behaviour can be explained by phenomenological Koops theory, in which dielectric materials have been described as two layer structure of Maxwell-Wagner type, where grains represent a well conducting layer, while grain boundaries represent a poorly conducting layer [43,44]. The grain possesses small value of dielectric constant and have dominant role at high frequencies whereas grain boundaries possess high value of dielectric constant and mainly influence the dielectric properties at low frequencies. In this model, the electrons reach the grain boundaries via hopping, but if these grain boundaries are highly resistive, they make electrons to get accumulated at the grain boundaries, and hence result in polarisation. With the rise in frequency of applied field, a decrease in polarization is observed because the electron start reversing their direction more frequently and hence the probability of electrons to reach the grain boundary decreases. This is reason why the ε′ and ε″ decrease with an increase in frequency. On the other hand, nearly constant loss (NCL) behaviour is observed, i.e. ε′ and ε″ seems to have very weak dependence on the frequency in the range from 1KHz to 1MHz [45]. This weak dependence often results from relaxations which involve highly localized motions of the charges moving in the asymmetric double-well potentials [43]. Figure 8 displays the variation of tan$\delta$ with frequency. The shifting of peak position toward the higher frequency side and the increase of peak height with increase in temperature indicates that these peaks are associated with thermally activated relaxation processes [46].

**3.7 Impedance Analysis**

The complex impedance spectroscopy measurements have been carried out to understand the effect of $NdFeO_3$ doping on the electrical characteristics of $Ba_{0.7}Sr_{0.3}TiO_3$ ceramics. In figure 9, the change in the real part of impedance (Z') within the frequency range (1Hz to 1MHz) and within the restricted temperature domain (583K-673K) has been displayed. The decreasing magnitude of Z' with increasing temperature confirm the typical NTCR behaviour. The convergent values of Z' in the high frequency region indicates towards the release of space charge and thus increase in *ac* conductivity with increasing temperature in the high frequency region and consequent reduction in the barrier properties of the material is observed [42]. The frequency dependence of $Z''(\omega)$ (in temperature range 583K-673K) illustrated in inset of figure 9 shows two dielectric relaxation peaks representing the bulk and grain boundary contribution. The broadening and shift in these peaks at high frequency region with temperature suggests that the electrical relaxation phenomenon in NFBST ceramics is temperature dependent. The merging of $Z''$ curves at all temperatures at the high frequencies is due to presence of space charges at the grain boundaries implying that the electrical responses are thermally activated. The dielectric relaxation time $\tau^z$ can be calculated for both grain and grain boundaries using the relation $\tau^Z = \frac{1}{\omega_{max}} = \frac{1}{2\pi f^Z_{max}}$ where $f^Z_{max}$ is the peak frequency of Z". Inset of Figure 9 shows the variation of relaxation time $\tau^Z$ as a function of temperature, which clearly predicts the decrease in relaxation time with temperature. The theoretical fit to the relaxation time $\tau^Z$ using Arrhenius relation $\tau^Z = \tau_0^Z\, e^{\frac{E^Z}{k_B T}}$, where $\tau_0^Z$ is the pre-exponential factor of relaxation time gives the activation energy. The activation energies $E^Z$ determined from the slope of linear fit in the Arrhenius plot for grains and grain boundaries are 0.58 eV and 1.63 eV respectively.

Figure 10 shows the complex impedance spectra (-Z' vs Z") and their fitting results at different temperatures (583 K- 673 K). Suppressed semicircular arcs with their centres below the real axis confirming non-Debye behaviour are observed. By using Z-VIEW software, the impedance data was best fitted with the equivalent circuit based on the brick-layer model as shown in figure 10. This circuit consists of an array of two sub-circuits i.e. $R_1 \parallel CPE_1$ and $R_2 \parallel CPE_2$; representing grain and grain boundaries effects respectively. Figure 11 clearly shows the variation of obtained fitting parameters R, C, and $\beta$ that the grain boundaries in NFBST are more resistive and capacitive than the grains. The activation energies ($E_a^{Rg} = 0.71 eV$ and $E_a^{Rgb} = 1.43 eV$) determined from linear fitting of resistance values (Rg & Rgb) to Arrhenius equation are in good agreement with the activation energies ($E_a^{\tau g} = 0.61 eV$ and $E_a^{\tau gb} = 1.57 eV$) obtained from the relaxation time ($\tau_g$ and $\tau_{gb}$) linear fit. From the nearly similar values of the activation energies one can estimate that the resistance and relaxation time are decreasing at the same rate [47].

In polycrystalline ceramics, reduction in electrical conductivity is mainly attributed to the interruption in the motion of charge carriers at the grain boundary. However, the traces of oxygen are lost by conductive grains during sintering at high temperature as per reaction [48,49]

$$O_o^x = \frac{1}{2} O_2(g) + V_o^x$$

$$V_o^x = V_o^{\blacksquare} + e'$$

$$V_o^x = V_o^{\blacksquare\blacksquare} + e''$$

and produce excess electrons and oxygen vacancies as follows (Written according to Kroger-Vink notation of defects)

$$\therefore \quad O_o^x = \frac{1}{2} O_2(g) + V_o^{\blacksquare\blacksquare} + 2e$$

These defects form barrier layers at the grain/grain-boundary interface and affect the impedance and capacitance of the system. Although, the re-oxidation taking place after sintering (during cooling) is limited to surface and grain boundaries, but it gives rise to a barrier by creating a difference between the resistance of the grain boundaries and the grains

[50]. Thus, from the above analysis it is estimated that the grain and grain boundary are separated by a potential barrier of about 0.44 eV (*i.e.* $E_a^{Rgb} - E_a^{Rg} = 1.43 - 0.71 eV = 0.72 eV$) forming a surface and internal barrier layer capacitor (IBLC).

### 3.8 Electric Modulus Analysis

To, get better insight of the electrical transport process occurring inside the material, and to distinguish the spectral components consisting of equivalent resistance and unequal capacitances, the dielectric response of non-conducting materials is made out via complex modulus analysis [51,52]. Figure 12 shows the variation of real part ($M'(\omega)$) and imaginary part ($M''(\omega)$) of the modulus $M^*$ at different temperatures. We can observe that the value of M' is very low (nearly zero) in low frequency regime (1Hz to 1 kHz) and it rises with frequency afterwards. The clearly seen dispersion of $M'(\omega)$ tending towards $M_\infty$ (the asymptotic value of $M'(\omega)$ at higher frequencies) may be attributed to the conduction phenomenon due to short range mobility of charge carriers which co-relates well to the scarce restoring force under the impact of induced electric field [51,53]. However, one can plausibly enunciate the long range mobility of charge carriers by observing the shift of dispersive region towards higher frequency side (>1 kHz). Further it is also noticed that behaviour of $M'(\omega)$ decreases with rise in temperature, thus confirms that short range mobility of charge carriers leads to conduction in the NFBST ceramics. From the Figure 12, the plot $M''(\omega)$ loss spectra clearly shows a peak at different temperatures in the higher frequency region i.e. centred at the dispersion region of the real part of electric modulus $M'(\omega)$. In the lower frequency region, the appeared loss peak maxima in the $M''(\omega)$ vs frequency spectra corroborate to long range mobile charge carrier region where these carriers migrates towards neighbouring sites following hopping conduction [54]. Therefore, the short range localized charge carriers get trapped in potential wells at the higher frequency side of peak maximum [43,55]. The possesses a low value of M" at lower frequencies i.e. from 1 Hz to 10 kHz at all the temperatures (583K-673K) not only confirms large value of capacitance in this region but also confirms the absence of electrode effect in the NFBST system. The theoretical fit to the experimental data using the Kohlrausch-William-Watts (KWW) function (Bergman modified) i.e. [56,57]

$$M'' = \frac{M''_{max}}{(1-\beta)+\left(\frac{\beta}{1+\beta}\right)\left[\beta\left(\frac{\omega_{max}}{\omega}\right)+\left(\frac{\omega}{\omega_{max}}\right)^\beta\right]} \quad (1)$$

gives the values of $M''_{max}$ and $\omega_{max}$ which are the maximum values of modulus M'' and angular frequency respectively. We can notice that the experimental data well fitted in this model (shown by red solid line). The mean dielectric relaxation time $\tau^M$ has been calculated using the relation $\tau^M = \frac{1}{\omega_{max}} = \frac{1}{2\pi f^M_{max}}$ where $f^M_{max}$ is peak frequency of M''. This relaxation time as a function of temperature is shown in the inset of figure 13. It is clearly observed that the relaxation time $\tau^M$ can be well fitted with Arrhenius relation $\tau^M = \tau^M_0 \, e^{\frac{E^M}{k_B T}}$ where $\tau^M_0$ is the pre-exponential factor of the relaxation time. The activation energy $E^M$ determined from the slope of the linear fit in the Arrhenius plot is 0.59 eV. The activation energy $E^Z$ corresponding to Z'' spectra represents the localized conduction (i.e. dielectric relaxation) where as that of $M''$ spectra i.e. $E^M$ attributes to the non-localized conduction (i.e. long range conductivity) [58]. The oxygen vacancies are the mobile charge carriers in perovskite ferroelectric materials which contribute in conduction and relaxation processes in the system [49,59]. With rise in temperature, these defects generate conduction electrons which play an essential role in conduction mechanism and their dependency can be understood via following reaction: [60]

$$V_O^{\bullet} \Rightarrow V_O^{\bullet\bullet} + e'$$

Usually, at low temperatures, singly ionized vacancies ($V_O^{\bullet}$) exists in ceramics. However, the rise in temperature excites the singly ionized vacancies ($V_O^{\bullet}$) to doubly ionized oxygen vacancies ($V_O^{\bullet\bullet}$) which thus creates the free charge carriers and consequently enhance the electrical conductivity.

Moreover, $M''/M''_{max}$ versus $log \, (f/f_{max})$ plot (figure 13) where $f_{max}$ is the loss peak frequency, indicates temperature dependent behaviour of the relaxation process in the NFBST system.

### 3.9 Electrical Conductivity Analysis

The transport behaviour of the charge carriers can be well elucidated by electrical conductivity as it not only depends on dielectric properties but also on the sample capacitance. The ac conductivity ($\sigma_{ac}$) can be calculated using a relation; $\sigma_{ac} = \omega \varepsilon \varepsilon_0 \, tan\delta$ where $\omega$ is the angular frequency and $\varepsilon_0$ is the permittivity in free space. The mobile charge carriers reach to a displacement state from its actual position through hopping conduction between two potential energy minima and contribute towards frequency dependant conductivity in relaxation phenomenon [61–64]. The aforementioned conduction behaviour follows the universal Jonscher`s power law, which is expressed as

$$\sigma_{ac}(\omega) = \sigma_0 + A\omega^n \qquad (2)$$

where $\sigma(\omega)$ is the total conductivity, $\sigma_0$ is the frequency independent or 'DC' part which is related to dc conductivity and the second term of constant phase element (CPE) type. The frequency exponent $n$ ($0 < n < 1$) is frequency independent, but the temperature and material dependant parameters represent the degree of interaction between mobile ions with lattices around them and A is the temperature dependent pre-exponential factor (constant) which determines the strength of polarization.

In general, the frequency dependence of conductivity does not follow the simple power relation as given above but follows a double power law given as

$$\sigma_{ac}(\omega) = \sigma_0 + A1\omega^{n1} + A2\omega^{n2} \qquad (3)$$

where $\sigma o$ is the frequency-independent (electronic or dc) part of ac conductivity. The exponent $n1$ ($0 \leq n1 \leq 1$) characterizes the low-frequency region, corresponding to translational ion hopping and the exponent $n2$ ($0 < n2 < 2$) characterizes the high-frequency region, indicating the existence of well localized relaxation / reorientational process, the activation energy of which is ascribed to reorientation ionic hopping. Further, it may be inferred that the slope $n1$ is associated with grain-boundary conductivity (with large capacitance values) whereas $n2$ depends on grain conductivity (with smaller capacitance values) [65]. Figure 14 shows the frequency-dependent conductivity spectra at selected temperatures (573K-648K) of NFBST. From the first glimpse, it can be clearly observed that there are two plateau regions (marked in figure 14) i.e. conductivity spectrum shows two frequency dependent dispersion regions. The low frequency plateau response is observed in region-I and frequency dispersion at higher frequency in region-II. The non linear curve fit to Johnscher`s double power law at different temperatures is shown in the figure 14 (represented by red solid line) whereas the inset of figure 14 shows the variation of $\sigma_{dc}$ with inverse of absolute temperature (1000/T). The DC conductivity shows almost a linear rise with temperature exhibiting the negative temperature coefficient of resistance (NTCR) behaviour. This variation of $\sigma_{dc}$ with temperature is explained by a thermally activated transport of Arrhenius type; $\sigma_{dc} = \sigma_0\, e^{\frac{E_a}{k_B T}}$ where $\sigma_0$, Ea and k$_B$ represent the pre-exponential term, the activation energy of the mobile charge carriers and Boltzmann's constant, respectively. The dc activation energy is estimated to be 0.59 eV which lies in the range of doubly ionized oxygen vacancy activation energy [12]. Hence, it is can be assumed that the doubly ionized oxygen vacancies are the main conduction carriers in this temperature range which act as polarons created by loss of oxygen from the lattice distortion [9].

Figure 15 shows the variation of fitting parameters 'n' and 'A'. It is observed that the value of $n_1$ and $n_2$ remains below 1 and it decreases with temperature which ensures that the ac conductivity in the sample arises because of Correlated Barrier Hopping (CBH) over the same barrier [52]. The transfer of electrons over the barrier having corresponding potential takes place via thermal activation which can be well understood by CBH model [66]. The enhancement in concentration of oxygen vacancies generated due to intrinsic or extrinsic processes are responsible for typical increase of the conductivity. However, the intrinsic oxygen vacancies may generate due to high temperature heat treatment of the samples and extrinsic oxygen vacancies may generate to maintain the charge neutrality.

## CONCLUSIONS:

The $NdFeO_3$ modified $Ba_{0.7}Sr_{0.3}TiO_3$ ceramics were prepared by a solid-state reaction route and characterized by various experimental techniques. The X-ray analysis using the Rietveld refinement revealed the existence of mixed phase i.e. both pseudocubic (84.18 %) and hexagonal (15.82 %) phase at room temperature. The deconvoluted Raman data also confirmed the existence of mixed phase in NFBST sample. The SEM image showed that doping of $NdFeO_3$ in BST results in dense microstructure with well packed grains of different sizes. The centrosymmetric crystal structure has resulted in lossy P-E loop where as for weak ferromagnetic character at room temperature, the ferromagnetic exchange mechanism can be held responsible. The dielectric response in NFBST ceramic is frequency dependent and thermally activated. Complex impedance spectroscopy have been studied to understand the behaviour of electrical parameters such as impedance (Z', Z"), modulus (M', M"), and conductivity as a function of both frequency and temperature. The complex impedance plots show both bulk and grain boundary contribution and it is found that NFBST ceramic exhibits NTCR behaviour as the grain and grain boundary resistance decreases with rise in temperature. The imaginary part of impedance ($Z''$) and modulus ($M''$) spectra show the distribution of relaxation times where as their scaling behaviour further suggest that the distribution is temperature independent. The detailed complex modulus and ac conductivity analysis revealed that the hopping mechanism can be responsible for electrical transport processes in NFBST. The ac conductivity spectrum follows obey Jonscher's universal power law and the parameters obtained from fitting of using double power law successfully justifies the CBH model. The estimated activation energies confirm that the ionized oxygen vacancies formed by loss of oxygen from crystal lattice at higher sintering temperatures contribute in the relaxation and conduction processes in NFBST ceramic.

**Acknowledgement:** One of the authors Anumeet Kaur would like to thank the University Grants Commission (UGC BSR) for fellowship. Authors are also thankful to Dr. A.M Awasthi from UGC-DAE Consortium for Scientific Research, Indore and IUAC New Delhi, India for providing research facilities.

**Table 1:** The Refined parameters calculated from Reitveld refinement of NFBST Sample

| Results of the Rietveld refinement for $(NdFeO_3)_{0.1}$ - $(Ba_{0.7}Sr_{0.3}TiO_3)_{0.9}$ | | | | | |
|---|---|---|---|---|---|
| **Tetragonal phase with space group *P4mm*,** | | | | | |
| **Cell parameters:- a = b = 3.95810 (10) Å, c = 3.95980 (6) Å Volume = 62.036 (0.002) (Å)³** | | | | | |
| **Ions** | $x_T$ | $y_T$ | $y_T$ | $B_{iso}(\text{Å}^2)$ | Occupancy |
| $Ba^{2+}$ (1a) | 0.00000 | 0.00000 | 0.00000 | 1.812 | 0.079 |
| $Sr^{2+}$ (1a) | 0.00000 | 0.00000 | 0.00000 | 1.812 | 0.034 |
| $Nd^{3+}$ (1a) | 0.00000 | 0.00000 | 0.00000 | 1.812 | 0.013 |
| $Ti^{4+}$ (1b) | 0.50000 | 0.50000 | 0.90055 | 1.143 | 0.112 |
| $Fe^{3+}$ (1b) | 0.50000 | 0.50000 | 0.90055 | 1.143 | 0.013 |
| $O_I^{2-}$ (1b) | 0.50000 | 0.50000 | -0.42621 | 0.543 | 0.125 |
| $O_I^{2-}$ (2c) | 0.50000 | 0.00000 | 0.86878 | 1.989 | 0.250 |
| **Hexagonal phase with space group *P63/mmc*** | | | | | |
| **Cell parameters:- a = b = 2.61375 (6) Å, c = 6.03904 (4) Å, Volume = 35.729 (1.227) (Å)³** | | | | | |
| **Ions** | $x_H$ | $y_H$ | $y_H$ | $B_{iso}(\text{Å}^2)$ | Occupancy |
| $Ba^{2+}$ (2b) | 0.00000 | 0.00000 | 0.25000 | 0.354 | 0.052 |
| $Ba^{2+}$ (4f) | 0.33330 | 0.66670 | 0.96100 | 1.139 | 0.105 |
| $Sr^{2+}$ (2b) | 0.00000 | 0.00000 | 0.25000 | 0.354 | 0.023 |
| $Sr^{2+}$ (4f) | 0.33330 | 0.66670 | 0.96100 | 1.139 | 0.045 |
| $Nd^{3+}$ (2b) | 0.00000 | 0.00000 | 0.25000 | 0.354 | 0.008 |
| $Nd^{3+}$ (4f) | 0.33330 | 0.66670 | 0.96100 | 1.139 | 0.017 |
| $Ti^{4+}$ (2a) | 0.00000 | 0.00000 | 0.25000 | 0.957 | 0.075 |
| $Ti^{4+}$ (4f) | 0.33330 | 0.66670 | 0.84633 | 0.877 | 0.150 |
| $Fe^{3+}$ (2a) | 0.00000 | 0.00000 | 0.00000 | 0.957 | 0.008 |
| $Fe^{3+}$ (4f) | 0.33330 | 0.66670 | 0.84633 | 0.877 | 0.017 |
| $O_I^{2-}$ (6h) | 0.51850 | 0.03700 | 0.25000 | 2.021 | 0.250 |
| $O_I^{2-}$ (12k) | 0.83490 | 0.66980 | 0.08020 | 1.061 | 0.500 |
| **%Molar** | | 84.18 % (Tetragonal) | | 15.82 % (Hexagonal) | |
| ***R*-factors** $R_{B(Tetragonal)} = 4.65$, $R_{B(Hexagonal)} = 2.24$ $R_p = 6.67$, $R_{w-p} = 8.82$, $R_{exp} = 7.12$, $\chi^2 = 1.53$ | | | | | |

**Figures**

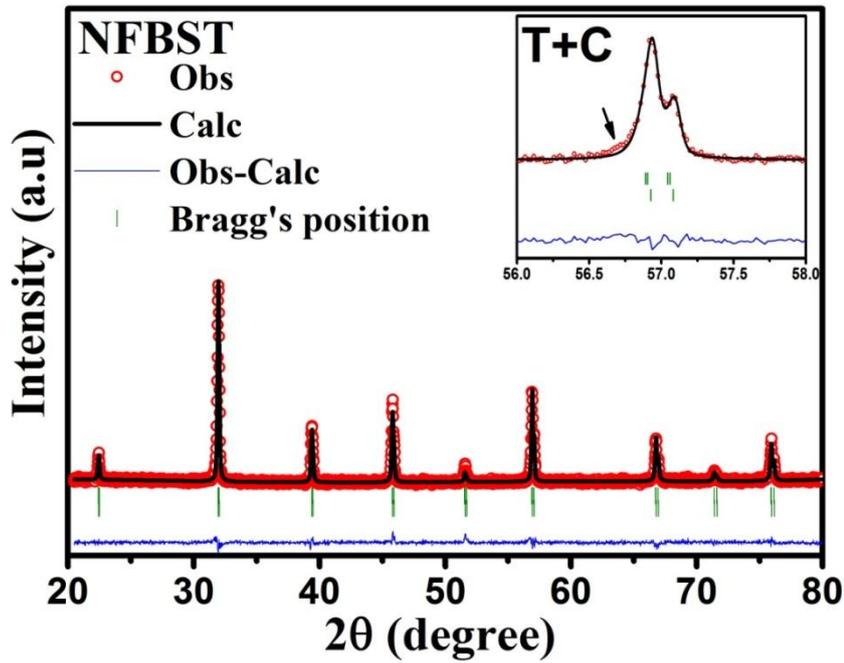

Figure 1(a): Observed and refined XRD pattern of NFBST Sample obtained using Tetragonal and Cubic Structure. Inset shows the mismatch between observed and refined data.

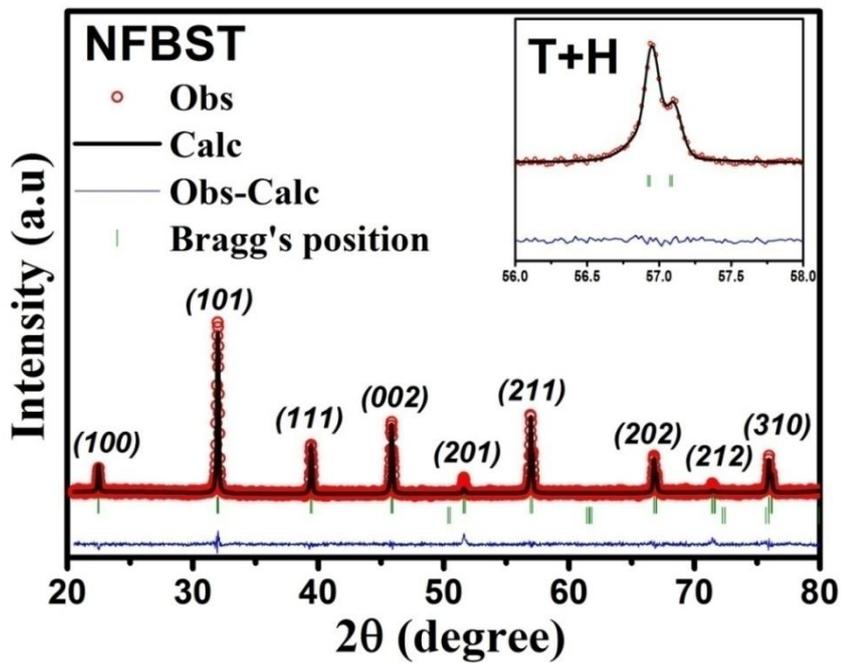

Figure 1(b): Observed and refined XRD pattern of NFBST Sample obtained using Tetragonal and Hexagonal Structure. Inset shows the perfect match of observed and refined data.

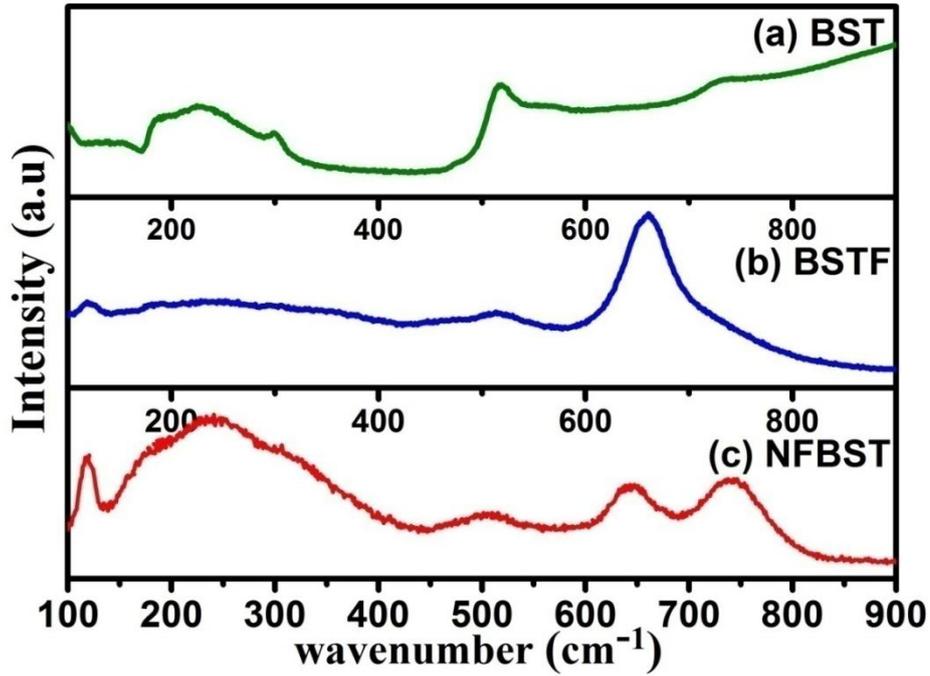

Figure 2(a): Room temperature Raman Spectra of (a) BST (b) BSTF and (c) NFBST sample

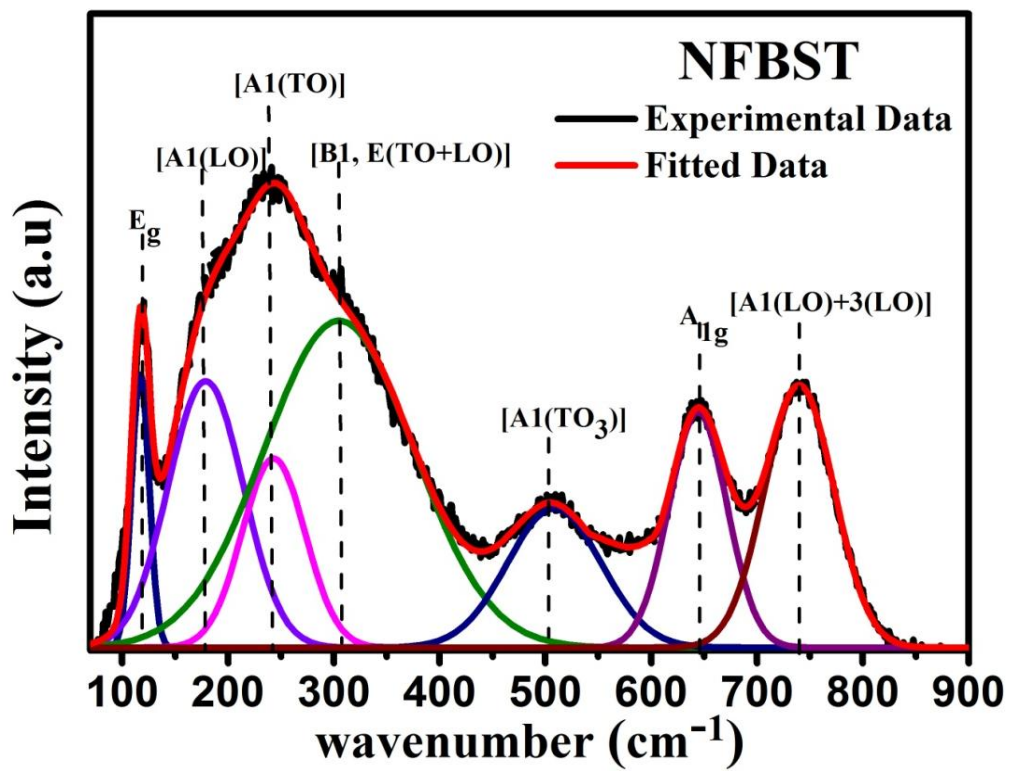

Figure 2 (b): Deconvoluted Raman Spectra of NFBST sample

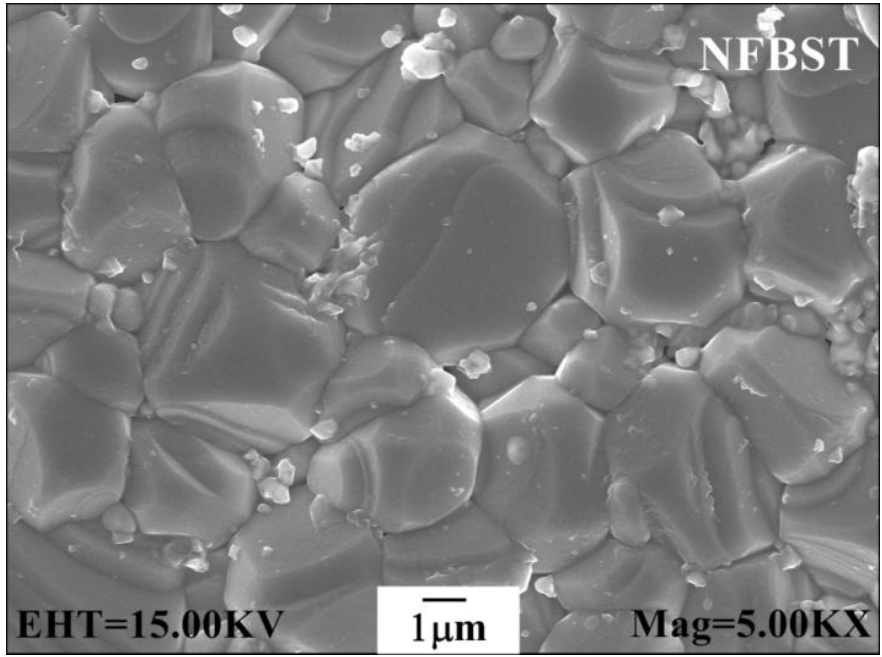

Figure 3: The SEM images of NFBST ceramics recorded at magnification of 5 KX.

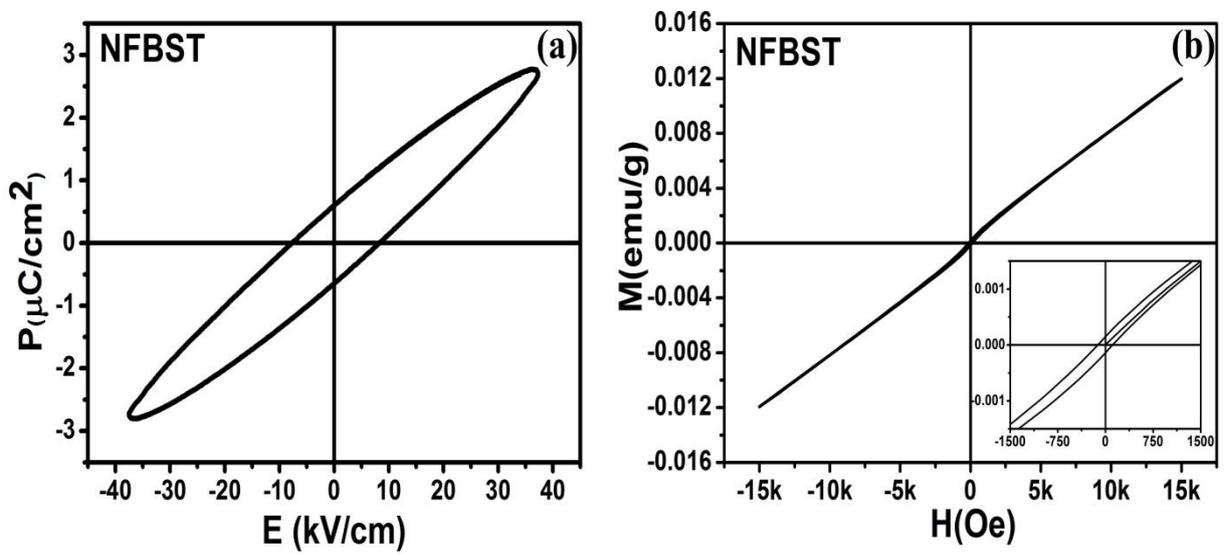

Figure 4: (a) Room temperature Ferroelectric loop (P-E) and (b) Room temperature Magnetic Behaviour of the NFBST sample

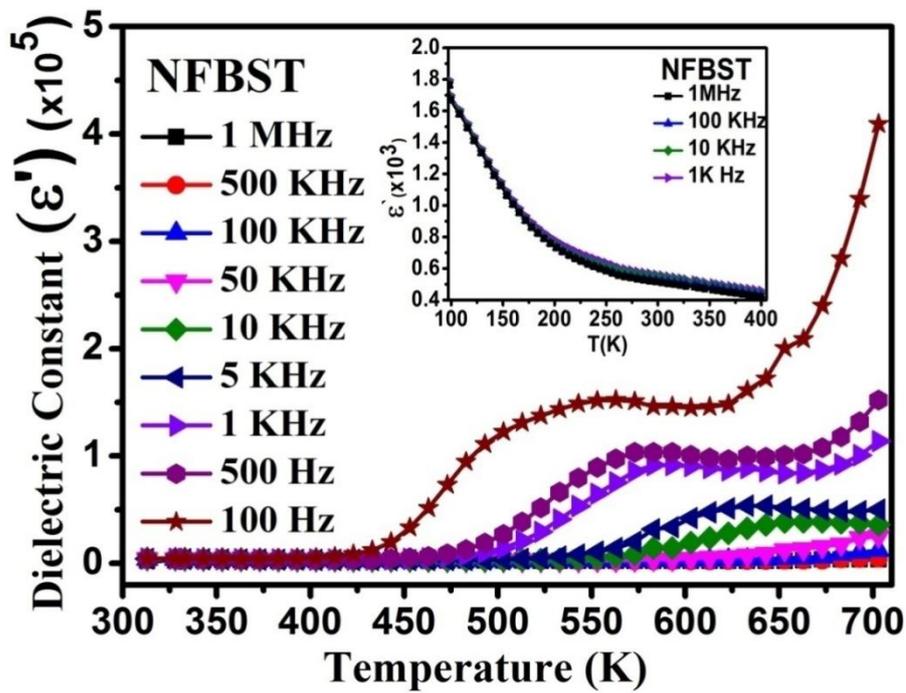

Figure 5. Temperature dependence of ε′ for NFBST sample in range from 300K to 700K. Inset shows variation of ε′ vs T in low temperature regime from 100K to 400K.

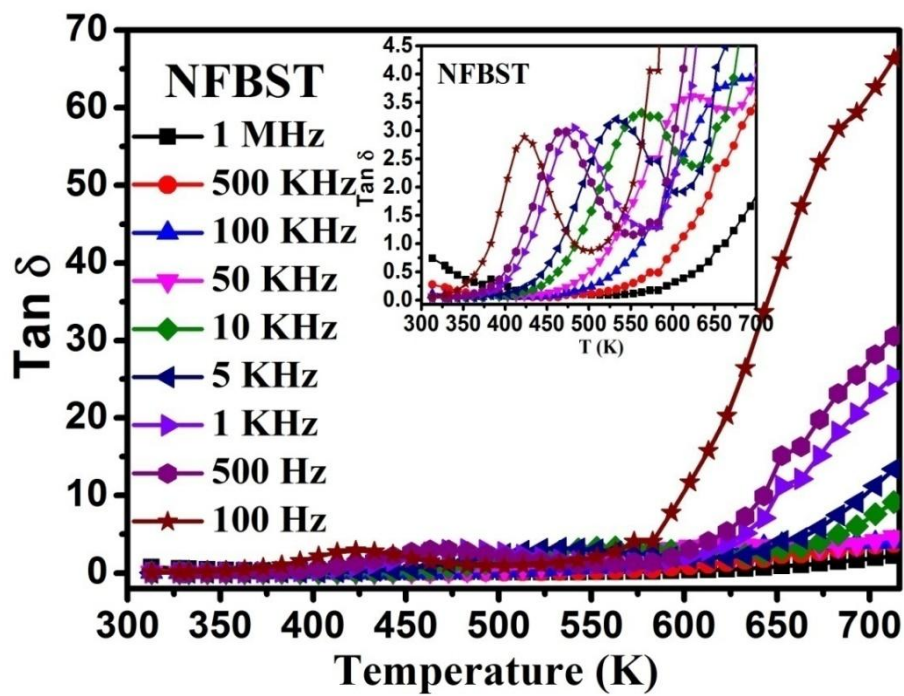

Figure 6. Temperature dependence of tan δ for NFBST sample in range from 300K to 700K. Inset shows enlarged view of variation.

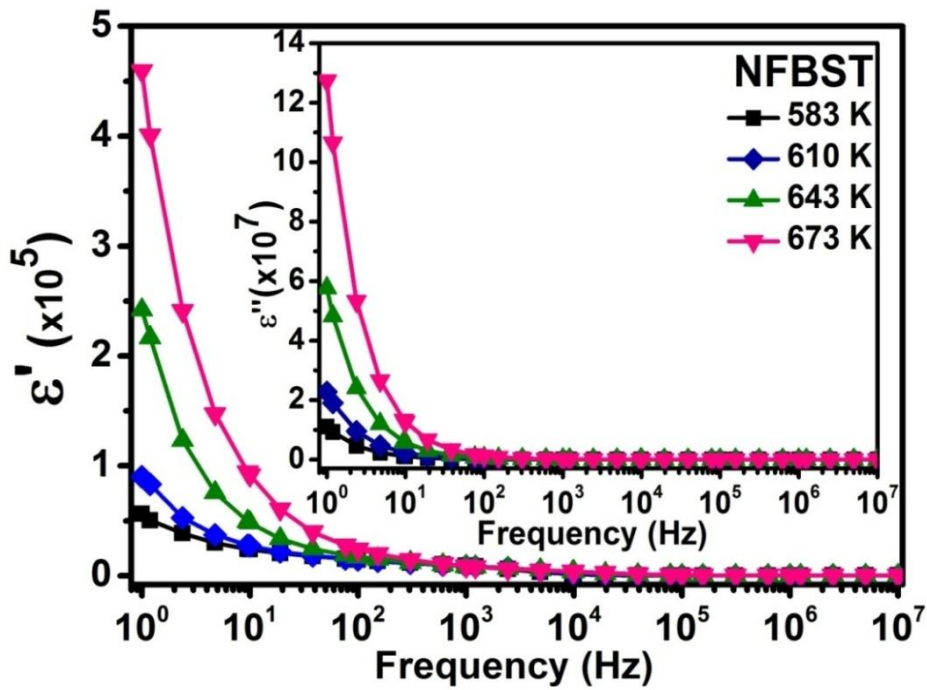

Figure 7. Frequency dependence of ε′ and ε″ with frequency for NFBST Sample

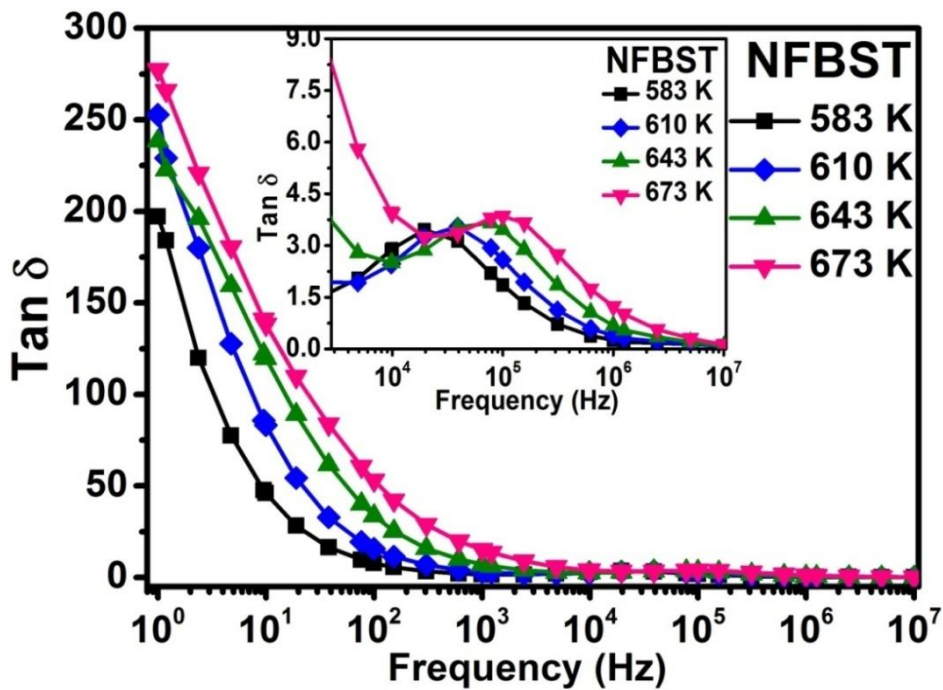

Figure 8. Frequency dependence of tan $\delta$ of NFBST Sample. Inset shows enlarged view of variation.

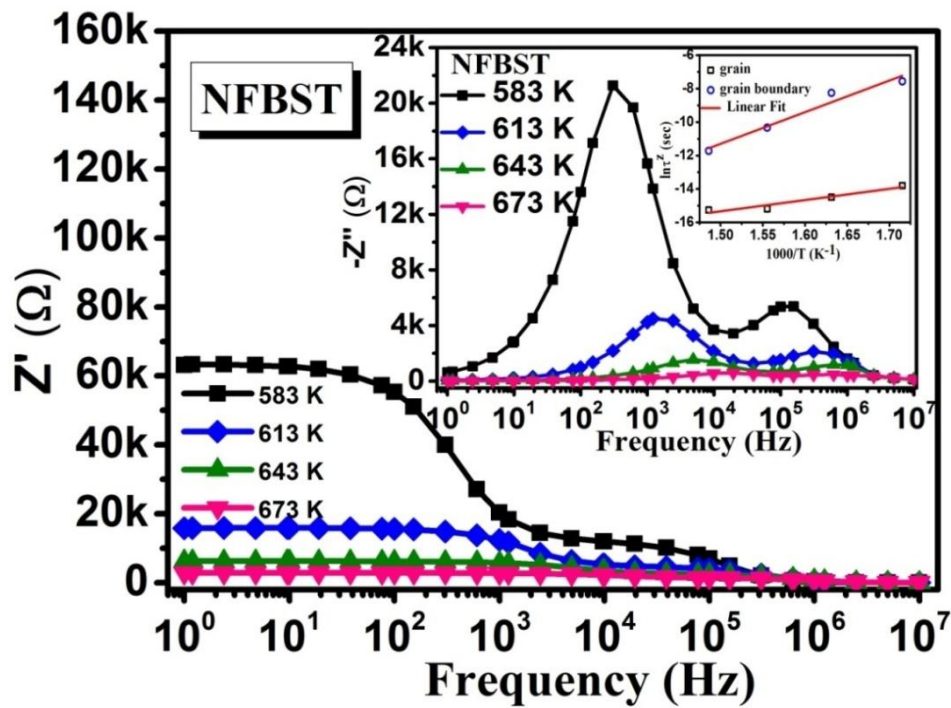

Figure 9. Frequency dependence Z′ for BSTF. Inset shows the Z″ versus Frequency variation and the Arrhenius fit to the relaxation time of grains and grain boundaries.

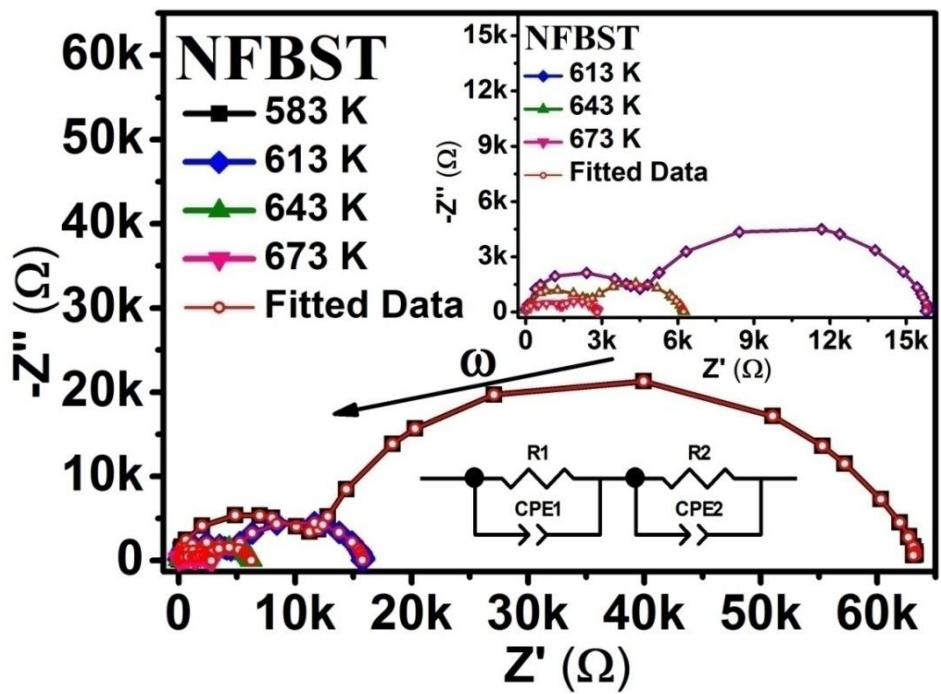

Figure 10. Cole-Cole plots of Z′ vs Z″ at different temperatures. Inset shows the enlarged view of data.

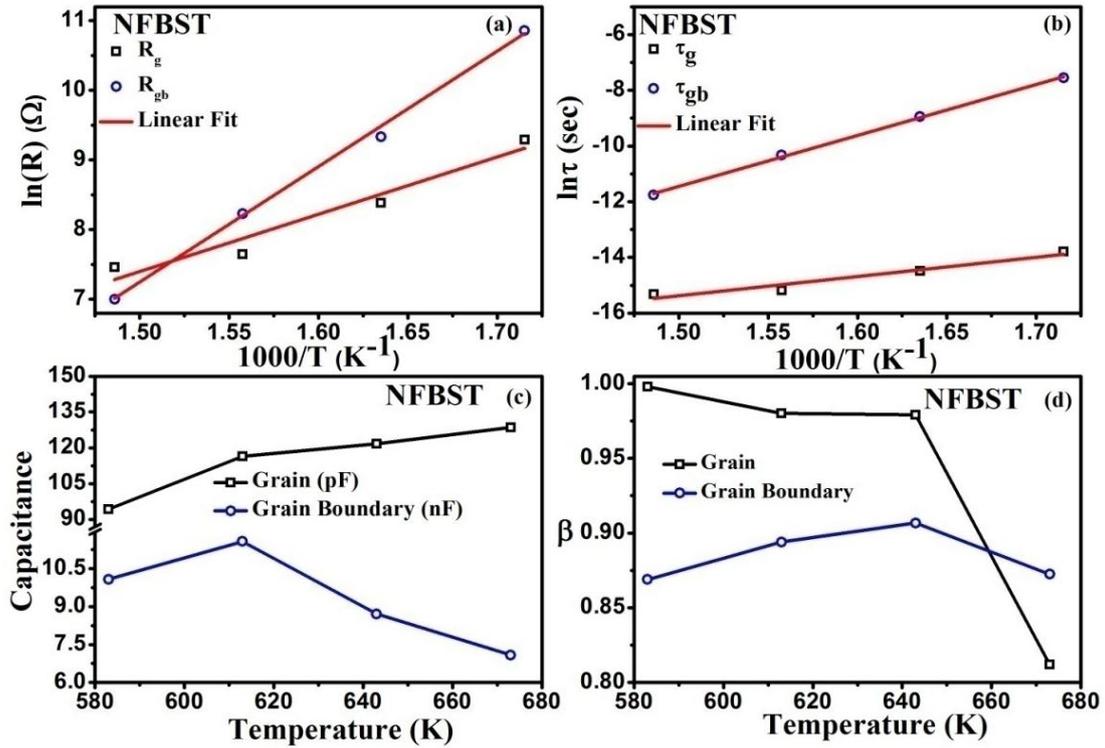

Figure 11. Arrhenius plots of (a) Resistance, (b) Relaxation time, (c) capacitance, and (d) β variation for grain and grain boundary obtained from the fitting of equivalent circuit at various temperatures.

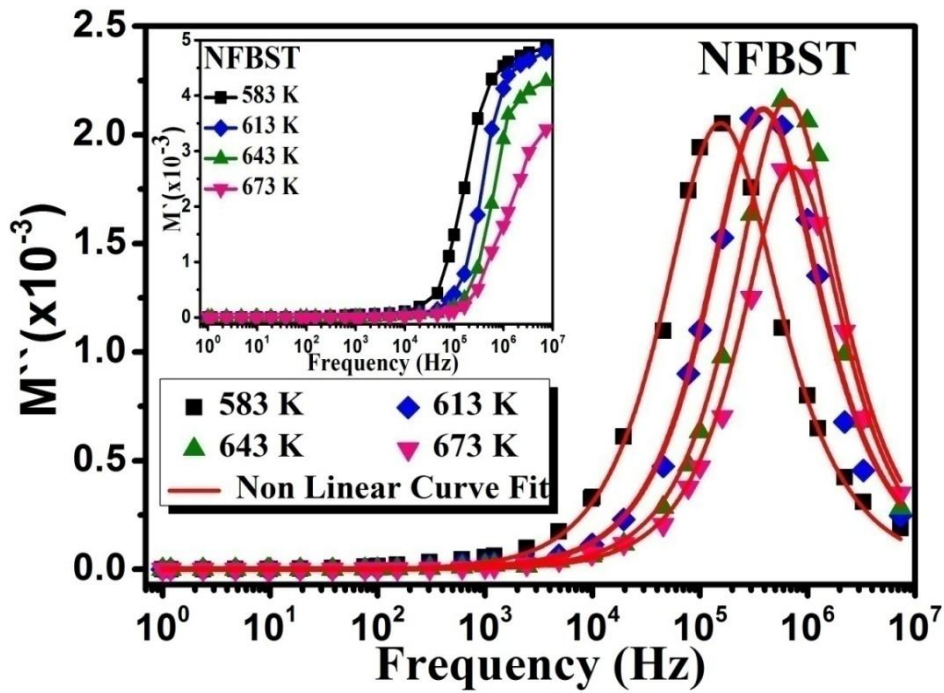

Figure 12. Frequency dependence of M″. Inset shows the M′ versus Frequency for NFBST

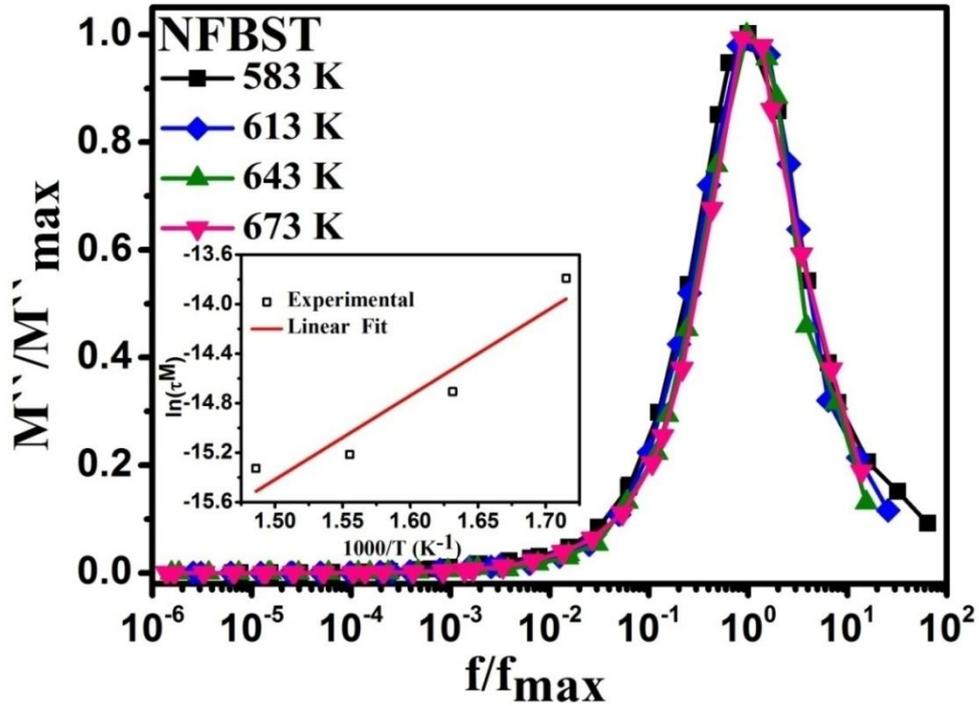

Figure 13. Scaling behaviour of imaginary part of modulus (M") spectra for NFBST. Inset shows the Arrhenius fit to the relaxation time.

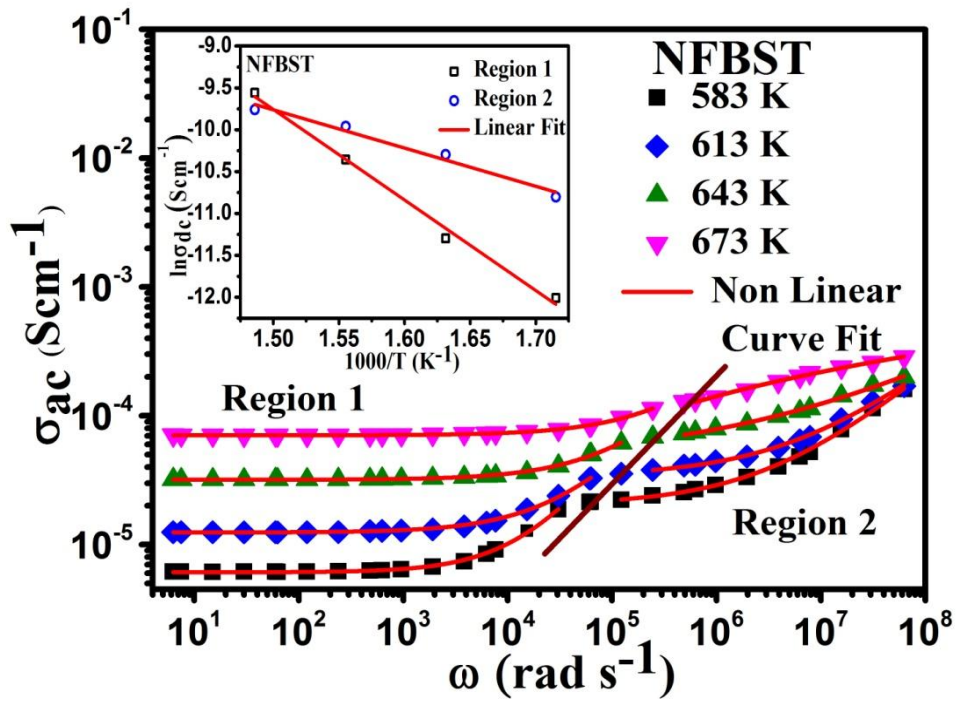

Figure 14. Frequency dependence of ac conductivity in NFBST. Inset shows the Arrhenius fit to the dc conductivity.

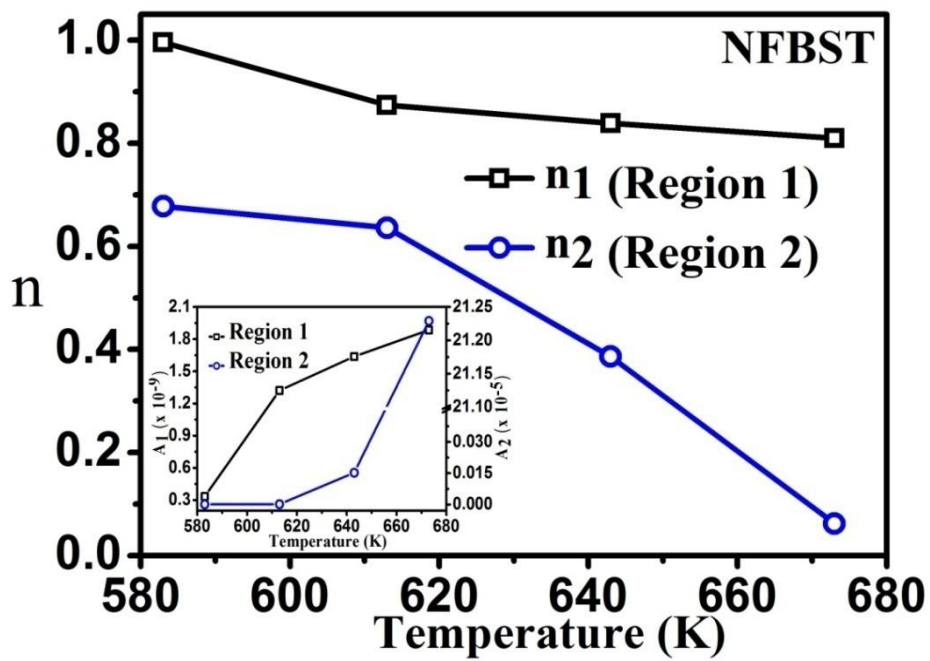

Figure 15. Variation of fitting parameters obtained from ac conductivity.